\documentclass[pre,twocolumn,nofootinbib]{revtex4-1}
\usepackage{graphicx}
\usepackage{amsmath}
\usepackage{bm}
\usepackage{amssymb}
\usepackage{tikz}
\usetikzlibrary{cd}
\usepackage{mathtools}

\begin{document}
\title{Good Fibrations: Packing Rules for Diabolic Domains}

\author{Randall D.~Kamien}
\affiliation{Department of Physics and Astronomy, University of Pennsylvania, 209 South 33rd Street, Philadelphia, Pennsylvania 19104, USA}
\affiliation{Department of Mathematics, University of Pennsylvania, 209 South 33rd Street, Philadelphia, Pennsylvania 19104, USA}
\author{Thomas Machon}
\affiliation{H.H. Wills Physics Laboratory, University of Bristol, Bristol BS8 1TL, United Kingdom}



\begin{abstract}
We describe a theory of packing hyperboloid `diabolic' domains in bend-free textures of liquid crystals.  The domains sew together continuously, providing a menagerie of bend-free textures akin to the packing of focal conic domains in smectic liquid crystals. We show how distinct domains may be related to each other by Lorentz transformations, and that this process may lower the elastic energy of the system. We discuss a number of phases that may be formed as a result, including splay-twist analogues of blue phases.  We also discuss how these diabolic domains may be subject to ``superluminal boosts'', yielding defects analogous to shocks waves. We explore the geometry of these textures, demonstrating their relation to Milnor fibrations of the Hopf link. Finally, we show how the theory of these domains is unified in four-dimensional space.

\end{abstract}

\maketitle
\begin{figure*}
\begin{center}
\includegraphics{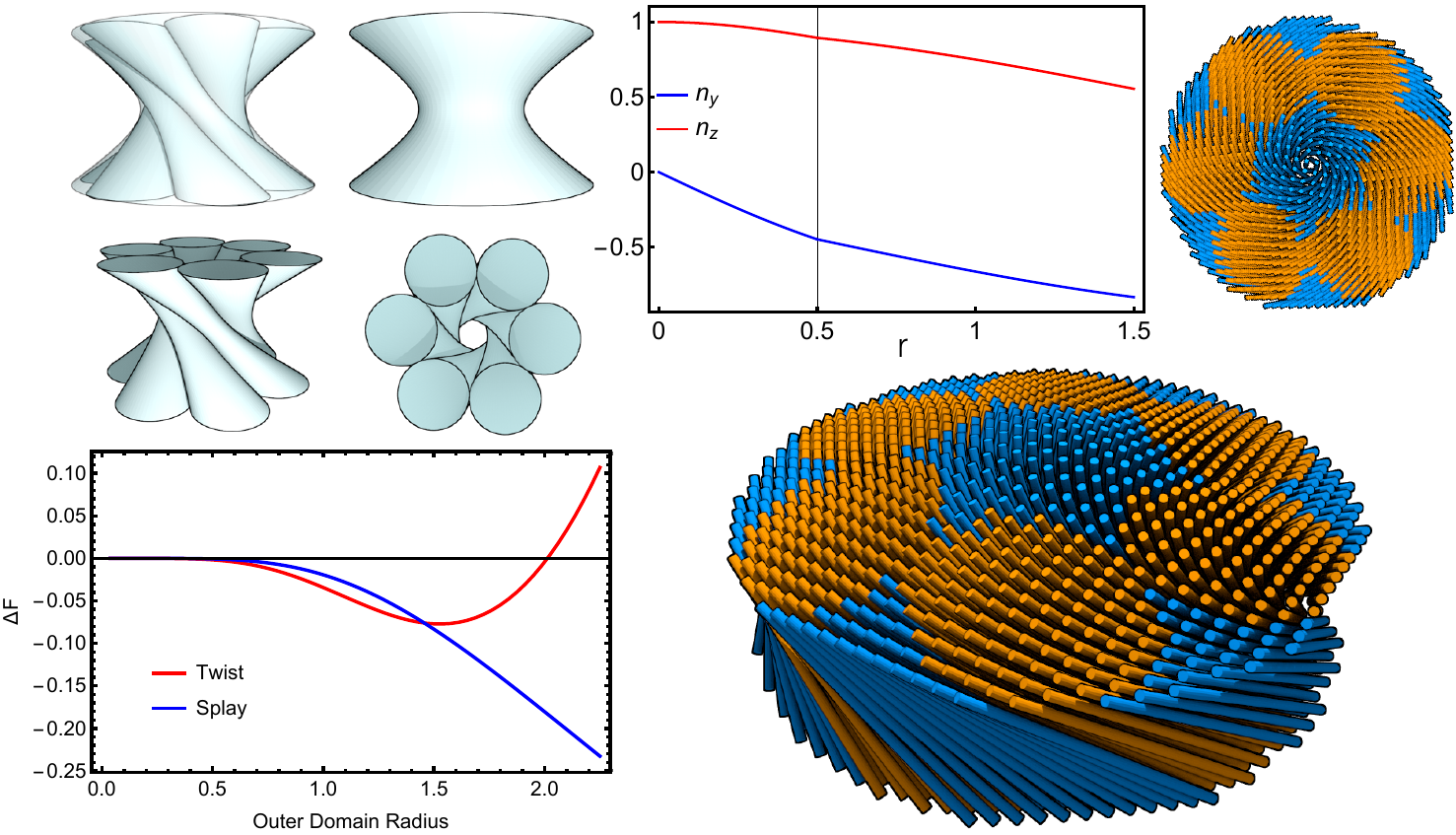}
\caption{Top left: hexagonal packing of skew domains inside an outer hyperboloid of radius $\rho$ and height $4$. Bottom left: change in splay and twist between a single large domain of radius $\rho$ and a hexagonal packing of subdomains inside a domain of radius $\rho$. The twist is favoured for an outer hyperboloid radius of $\rho = \pi/2$, which likely sets the size of the domain. Top center: graph showing the $n_z$ and $n_y$ components along the line $y=z=0$, where the discontinuity in the derivative can be seen at the domain boundary. Bottom right and top right: integral curves of the director field for the hexagonal configuration, drawn as cylinders. The gold cylinders are inside boosted subdomains, the blue cylinders are from the original, unboosted, texture.}
\label{fig:energy}
\end{center}
\end{figure*}

The Ising model of a ferromagnet captures the essence of long-range order.  Below a critical coupling the spins globally align along a common direction.  However, domain walls, across which the spin jumps from up to down, typically proliferate real samples.  The failure to create a single monodomain can arise from the kinetics of nucleation but can also arise from global energy minimization once the bulk magnetic field is added to the model \cite{LanLif}.  Ordered systems of all sorts can be split along this dichotomy. Unstressed crystals contain domains which, over time, coarsen on their way to a single crystal while crystals under stress will adopt dislocation-ridden ground states in order to globally minimize the strain energy \cite{hirth1992theory}.  A closed flux line in a type II superconductor 
or a vortex ring in a superfluid will eventually shrink and vanish while, in the presence of a magnetic field or a rotating container the two systems will find ground states with coexisting normal and ``super'' components with magnetic field or vorticity persisting in the normal regions \cite{Ab,HV1,HV2}.  Liquid crystalline materials, soft as they are, enjoy these equilibrium constructs as well.  The Renn-Lubensky twist-grain-boundary phase \cite{Renn:1988p2132} in smectic liquid crystals completed de Gennes' analogy between the smectic and the superconductor \cite{dG}.    These examples, however, all require the introduction of {\sl topological defects} whether they be $\pi$-walls, disclinations, or dislocations \cite{plw}.  However, smectic liqud crystals reveal an even more fragile construction based upon focal conic domains (FCDs).  Revealed by their tell-tale conic sections visible in bright-field, these domains are composed of equally-spaced smectic layers which match continuously across the domain walls, as in martensitic domains in crystals.  Between different FCDs the remaining space can be filled with other equally-spaced smectic layers giving rise to Friedel's law of corresponding cones \cite{friedel}, bend walls \cite{grainb}, and complex ``flower'' patterns \cite{flower1,flower2}.  Here, we describe a still softer way of cutting and sewing regions together based on {\sl diabolic domains}: these are diabolo-shaped liquid crystalline domains with zero bend as discussed in \cite{chaturvedi2020gnomonious}.  Multiple domains can be connected through bend-free textures.  In doing so, it is possible to lower the overall Frank free energy of, for instance, chiral materials that would form blue phases but with very large bend moduli.  As shown in Fig.~\ref{fig:energy} a single diabolic domain breaks up into smaller ones, lowering the splay and twist energies of the outer domains by replacing the interior texture with the center of a different bend-free texture.  This allows for a more uniform twist, favorable for a chiral nematic.  This conquer and divide approach can be repeated until the energy cost of the domain walls becomes prohibitive.  

In the following, energetic considerations will be calculated via the Frank free energy of liquid crystals~\cite{frank58},
\begin{eqnarray}
F&=& \frac{1}{2}\int {\rm d} V \, \Bigg\{ K_1[{\bf n}(\nabla \cdot {\bf n})]^2  + K_2[{\bf n} \cdot (\nabla \times {\bf n}) + q_0 ]^2  \\ &&\quad+ K_3 [({\bf n} \cdot \nabla) {\bf n} ]^2 + K_{24} \nabla \cdot [ {\bf n} (\nabla \cdot {\bf n}) - ({\bf n} \cdot \nabla) {\bf n}]\Bigg\} \nonumber,
\end{eqnarray}
comprising splay, twist, bend, and saddle-splay respectively.  Herein, we consider a system where $K_3$ is large, and study the limit where the bend distortions vanish, $({\bf n} \cdot \nabla) {\bf n} = 0$. Bend measures the (geodesic) curvature of the integral curves of the director field; zero bend implies the director field traces out a set of straight lines in the sample.

Recently a chiral bend-free texture of nematic liquid crystals, rotationally symmetric around the $z$-axis was considered as a potential building block for a splay-twist phase \cite{chaturvedi2020gnomonious}
\begin{equation} \label{ugh}
{\bf v}_0 = \left [x z+y, yz -x, 1+z^2 \right ], \quad {\bf n}_0 = {\bf v}_0 / |{\bf v}_0|.
 \end{equation}
${\bf n}_0$ can be thought of as a ``bend-free'' double-twist cylinder\footnote{As a matter of {\sl entente} we have reversed the handedness of this texture from right- to left-handed compared to \cite{chaturvedi2020gnomonious}.}.  This texture  ${\bf n}_0$ corresponds to the homogeneous double-twist texture associated to the Hopf fibration on the curved space $\mathbb{S}^3$ \cite{sethna1983relieving}.  However, unlike the more familiar stereographic projection of the Hopf fibration leading to linked circles in $\mathbb{R}^3$, here ${\bf n}_0$ arises from the {\sl gnomonic} projection of $\mathbb{S}^3$ onto $\mathbb{R}^3$ which, by construction, maps great circles to straight lines. We note that the homogeneous double-twist arrangement on $\mathbb{S}^3$ provided by the Hopf fibration is only one of a multitude of homogeneous nematic textures on three-dimensional manifolds \cite{new4,new3,new5}.  Whether the geometrical structures we describe in the following can be adapted to construct packings of frustrated textures of other kinds remains to be seen.

The integral curves of ${\bf n}_0$ lie on a family of constant tilt hyperboloids ${\bf n}_0\cdot\hat z = 1/\sqrt{1+\rho^2}$ where $\rho$ is the radius at the ``waist'' of the hyperboloid on the $z=0$ plane: $\rho^{-2}(x^2+y^2) - z^2 = 1$.  Each value of $\rho$ gives a different hyperboloid and, together, this collection of hyperboloids fills $\mathbb{R}^3$.  We have a foliation of space by hyperboloids with a singular hyperboloid along the $z$-axis, as shown in Figure~\ref{fig:standard} (in $\mathbb{S}^3$ these correspond to concentric tori).  Now consider a particular hyperboloid with radius $\rho_0$.  The hyperboloid is invariant under
 \begin{equation}\label{lt}
 x'=\gamma(x-\beta \rho_0 z),\quad  z'=\gamma\left (z-\frac{\beta}{\rho_0}x\right),
 \end{equation}
where $\beta=v/\rho_0$, $\gamma^2(1-\beta^2)=1$, and $v$ is an arbitrary parameter: the Lorentz transformation treating $z$ as time with the ``speed of light'' set to $\rho_0$. For reasons that will become clear, we refer to this as the ``little Lorentz'' transformation, and the group $\rm{SO}(2,1)$ of such transformations as the little Lorentz group\footnote{Throughout we use ${\rm SO}(p,q)$ to refer to the proper, orthochronous ({\sl i.e.}, restricted) group  ${\rm SO}^+(p,q)$.}. 

Since the hyperboloid is ruled and the rules define the nematic texture, the invariance of the hyperboloid implies that the nematic texture is unchanged {\sl on the hyperboloid} with radius $\rho_0$.  However, consider an inner hyperboloid, centered around the $\hat z$-axis, with radius $\rho_1<\rho_0$.  Under the little Lorentz transformation two things happen: its cross section in the $z=0$ plane squashes the circle of radius $\rho_1$ into an ellipse with major axis $\rho_1$ and minor axis $\gamma^{-1}\rho_1$ (the Fitzgerald contraction), and the original centerline tilts from $(x,y,z)=(0,0,z)$ to $(x,y,z)=(v z,0,z)$.  Each hyperboloid is a ruled surface and so the ruling straight lines define a new, bend-free nematic texture that matches the nematic texture on the original $\rho_0$ hyperboloid. Applying the boost \eqref{lt} to ${\bf v}_0$ yields 
\begin{equation}\label{texture0}
{\bf v}_1 = \gamma {\bf v}_0 + \frac{\gamma \beta}{\rho_0} \left [ \rho_0^2-x^2, z \rho_0^2 - x y, y - x z \right ],
\end{equation}
normalizing ${\bf v}_1$ gives a new bend-free director field. The original texture ${\bf n}_0$ is skew -- no two integral curves are parallel -- and this property is preserved under the little Lorentz transformations. We therefore call textures obtained from \eqref{ugh} `skew textures', or `skew domains' when restricted to the interior of a hyperboloid.  

We could then construct a hybrid texture starting with ${\bf v}_0$ by cutting out the interior of the $\rho_0$ hyperboloid and replacing it with ${\bf v}_1$. Because the two textures agree on the $\rho_0$ hyperboloid, this yields a continuous director field -- no defects are created -- akin to the construction of focal conic domains in smectics, for instance. Does this hybrid configuration lower the Frank free energy? For this particular case the free energy is not lower. The original radially symmetric ${\bf v}_0$ has a lower free energy that ${\bf v}_1$. However, the concentric $\rho_0$ hyperboloids are not the only hyperboloids in \eqref{ugh}. As we show below, the texture \eqref{ugh} contains a huge number of `hidden' hyperboloids. These additional hyperboloids are not centered around the origin; they are shifted, tilted and sheared, an example is shown in Figure~\ref{fig:cnp}. We will demonstrate that the `drill and fill' construction can be applied to these hidden hyperboloids, where it does lower the free energy.  

\section*{Hidden hyperboloids in skew domains}
\begin{figure}
\begin{center}
\includegraphics[scale=1]{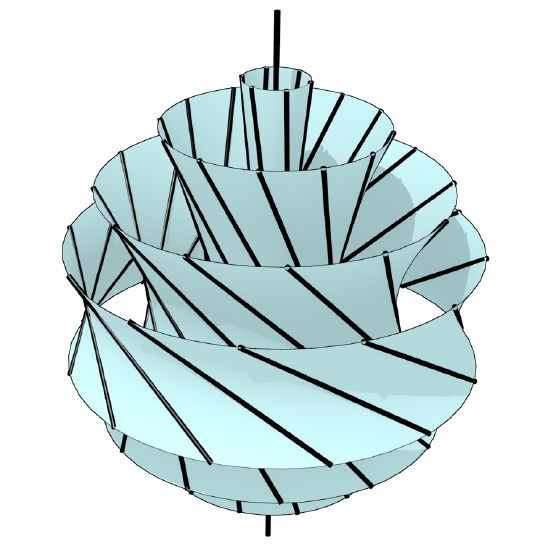}
\caption{Concentric hyperboloids in the texture \eqref{ugh}, given by \eqref{aform} with $a_0=(1-\rho_0^2)/2$ and $a_3=(1+\rho_0^2)/2$, $\rho_0$ the waist radius.}
\label{fig:standard}
\end{center}
\end{figure}

How do we find the `hidden' hyperboloids in ${\bf n}_0$? On $\mathbb{R}^3$ the structure treats the $z$-axis specially, but on the curved space $\mathbb{S}^3$, ${\bf n}_0$ corresponds to the Hopf fibration -- a homogeneous structure. Since there is no special point on $\mathbb{S}^3$, the concentric hyperboloids must also not be special, and we must be able to define a whole family of equivalent hyperboloid surfaces. 

To proceed we recall gnomonic projection and the Hopf fibration. Let $(x,y,z,w)$ be coordinates for $\mathbb{R}^4$, we take $\mathbb{R}^3$ to be the $w=1$ subspace $\mathcal{V}^3 \cong \mathbb{R}^3$, corresponding to $(x,y,z,1) \in \mathbb{R}^4$. Taking advantage of the fact that four real dimensions can be identified with two complex dimensions, we write $z_1 = x+ i y$, and $z_2 = z+ iw$, so that $ (z_1,z_2) \in \mathbb{C}^2$. The two real equations ${z_1} = \alpha {z_2}$
define a plane through the origin in $\mathbb{R}^4$ for each complex $\alpha$. We allow $\alpha = \infty$, so that $\alpha \in \mathbb{C} \cup \infty = \mathbb{C}{\bf P}^1\cong \mathbb{S}^2$, the Riemann sphere. Consider the intersection of the planes with the 3-sphere $\mathbb{S}^3$, given by $x^2+y^2+z^2+w^2 = |z_1|^2+|z_2|^2=1$. Each plane intersects $\mathbb{S}^3$ on a great circle, and the collection of great circles for all $\alpha \in \mathbb{C}{\bf P}^1$ fills up $\mathbb{S}^3$. This is the Hopf fibration, it can be thought of as a map $\mathbb{S}^3 \to \mathbb{S}^2$, with $(z_1,z_2) \mapsto \alpha \in \mathbb{C}{\bf P}^1 \cong \mathbb{S}^2$. Now  consider the intersection of the planes with the 3-space $\mathcal{V}^3$. For each $\alpha$, the corresponding plane intersects $\mathcal{V}^3$ along a line, giving a congruence of straight lines in $\mathbb{R}^3$. These lines are the integral curves of ${\bf n}_0$.   But this is the result of gnomonic projection: the northern hemisphere ($w>0$) of $\mathbb{S}^3$ may be identified with $\mathcal{V}^3 \cong \mathbb{R}^3$ via  $(x,y,z,w) \mapsto (x/w,y/w,z/w,1)$. Since this map preserves the value of $\alpha$, the integral curves of ${\bf n}_0$ are the gnomonic projection of the great circle fibers of the Hopf fibration. On $\mathbb{S}^3$, any two great circles on the three sphere that do not intersect must link. The linking of the circles in $\mathbb{S}^3$ leads, gnomonically, to a family of straight lines in $\mathbb{R}^3$ that are all {\sl skew} -- no two lines are parallel. Finally, note that as a director field, ${\bf n}_0$ arises as gnomonic projection of the complex vector field ${\bf h} = (i z_1, i z_2)$, the double twist texture on $\mathbb{S}^3$.

\begin{figure}
\begin{center}
\includegraphics[scale=1]{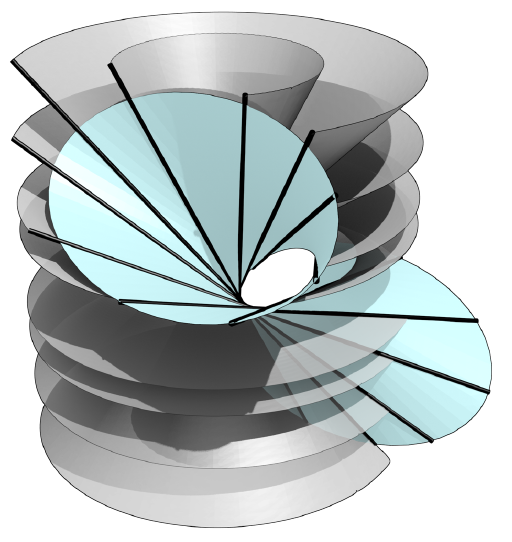}
\caption{Hidden hyperboloid in the texture \eqref{ugh}. The vertically-oriented surfaces are the family shown in Figure~\ref{fig:standard}. The black lines correspond to integral curves of ${\bf n}_0$, and intersect all hyperboloids on straight lines.}
\label{fig:cnp}
\end{center}
\end{figure}

In constructing ${\bf n}_0$, we chose the copy of $\mathbb{R}^3$ corresponding to $w=1$. But we could have projected to {\sl any} copy of $\mathbb{R}^3$ tangent to $\mathbb{S}^3$.  In some other projection the family of concentric hyperboloids in Figure~\ref{fig:standard} are no longer centered around the $\hat z$ axis, so hidden in ${\bf n}_0$ there must be more hyperboloids. On $\mathbb{S}^3 \subset \mathbb{C}^2$ the great circles of the Hopf fibration are traced out by a phase $(z_1, z_2) \to e^{i \phi }(z_1,z_2)$, that preserves the value of $\alpha$. The following quadratic expressions are invariant under that phase: $\vert z_1\vert^2 + \vert z_2\vert^2$, $\vert z_1\vert^2 - \vert z_2\vert^2$, $\hbox{Re}\, z_1{\bar z}_2$, $\hbox{Im}\, z_1{\bar z}_2$. Thus the general, real quadratic polynomial in $\mathbb{C}^2$ (with $a_\mu\in\mathbb{R}^4$) that contains the origin is
 \begin{eqnarray} \label{poly}
 0&=&a_0 \left[ \vert z_1\vert^2 + \vert z_2\vert^2 \right]+ a_1\left[ z_1{\bar z}_2 + z_2{\bar z}_1\right] \\ &&\quad+ a_2\left[ i z_1{\bar z}_2 -i z_2{\bar z}_1\right] + a_3\left[ \vert z_1\vert^2 - \vert z_2\vert^2 \right].\nonumber
  \end{eqnarray}
Note that there are no linear polynomials with this phase invariance and, because we will want to gnomonically project, non-zero constants are excluded.  Indeed, the homogeneity implies that if we consider solutions of the form $z_1=\alpha_0 z_2$ then \eqref{poly}) only depends on $\alpha_0$: the polynomial intersects entire planes in $\mathbb{C}^2$ and so therefore if a great circle intersects \eqref{poly} it lies entirely on it.  Moreover, under gnomonic projection that {\sl same} plane intersects $\mathcal{V}^3$ in a straight line and so \eqref{poly} is a quadratic surface ruled by the integral curves of the bend-free nematic texture \eqref{ugh}:
\begin{eqnarray}\label{aform}
0&=&(a_0+a_3)(x^2 + y^2) + 2a_1 (y + xz)  \\&&\quad +2a_2 (-x +yz) +(a_0-a_3) (1+z^2).\nonumber
 \end{eqnarray} 
When $a_1=a_2=0$, $a_0=(1-\rho_0^2)/2$, and $a_3=(1+\rho_0^2)/2$ we recover the symmetric hyperboloids shown in Figure~\ref{fig:standard}.  However, if $a_1$ or $a_2$ are non-zero we get tilted hyperboloids with elliptic cross section.  It is amusing, however, to note all these hyperboloids intersect the $z=0$ plane in circles, just not around the origin.  The shifts, tilts, and shears are the consequence of the gnomonic projection.  The hidden nested hyperboloids of the the texture \eqref{ugh} are revealed, as shown in Figure~\ref{fig:cnp}.

In matrix notation, \eqref{poly} is
 \begin{equation}\label{qform}
{\bf Z}^\dagger {\bf A} {\bf Z}\equiv \left[\begin{matrix}{\bar z}_1 & {\bar z}_2\end{matrix}\right] \left[\begin{matrix} a_0 + a_3 & a_1 - i a_2 \\a_1 + i a_2 & a_0 - a_3\end{matrix}\right]\left[\begin{matrix} z_1 \\ z_2\end{matrix}\right] =0,
 \end{equation}
where ${\bf A}$ is the defined Hermitian matrix and the column vector ${\bf Z}=[z_1,z_2]\in\mathbb{C}^2$.  But this quadratic form can only have non-vanishing solutions $\bf Z$ if ${\bf A}$ is indefinite. Thus its determinant must be negative, implying that the quadratic surface is a hyperboloid, and that $a_1^2+a_2^2+a_3^2 - a_0^2 >0$.  It is now natural to consider $a_\mu$ as a four-vector in (3+1) Minkowski space.  We will refer to transformations that preserve the space-time interval as ``Big Lorentz transformations.''  It is a distinct group from the little Lorentz group (in fact, the little Lorentz group is not a subgroup of the Big Lorentz group).  The four-vector $a_\mu$ is spacelike (and by homogeneity we could choose $a_\mu a^\mu=1$ so the magnitude is unimportant).   Since the Big Lorentz group ${\rm SO}(3,1)$ preserves the interval we find a family of quadratic surfaces in the gnomonic projection related by Big boosts.

The shifts, tilts and shears of the hyperboloids in \eqref{aform} can be understood in terms of the Big Lorentz transformation properties of $a_\mu$.  The quadratic form \eqref{qform} is invariant under 
${\bf Z}\rightarrow {\bf M}{\bf Z}$, and ${\bf A}\rightarrow [{\bf M}^\dagger]^{-1}{\bf A}{\bf M}^{-1}$ where ${\bf M}\in {\rm SL}(2,\mathbb{C})$.  Recalling that ${\rm SL}(2,\mathbb{C}) \cong {\rm SO}(3,1)$, and that it acts through conjugation on the quaternionic form in ${\bf A}$, we note that for any spacelike choice of $a_\mu$ we can perform a Big Lorentz transform to put it into a form with $a_0=a_1=a_2=0$ and $a_3=1$, the $\rho_0=1$ hyperboloid.  The action on $\bf Z$ by $\bf M$ generates a coordinate transformation on $(x,y,z,w)$ and so, in this new coordinate system we would see concentric, circular hyperboloids.  But the ${\rm SL}(2,\mathbb{C})$ transformation on $\bf Z$ gives rise to the famous M\"obius transformations of the Riemann sphere by acting on $z_1/z_2 \in \mathbb{C}{\bf P}^1$.   On the $z=0$ plane in $\mathcal{V}^3$, $z_2=i$ and we see that the Big boosts transform circles to circles.  Because ${\rm SL}(2,\mathbb{C})$ acts transitively, we deduce that any circle in the $xy$ plane is the (tilted) cross section of some hyperboloid hidden in \eqref{ugh}.  The ${\rm SO}(3,1)$ boosts of $a_\mu$ shift the hyperboloids \eqref{aform} but leave the texture \eqref{ugh} invariant -- they simply relabel the fibers of the Hopf fibration via M\"obius transformations of the Riemann sphere.

\section*{Diabolo domains}
\label{sec:packing}

We can now return to the original quest: start with ${\bf n}_0$ and choose any circle in the $xy$-plane.  That defines a tilted, elliptical hyperboloid $H$ through the integral curves of the nematic field.  But $H$ has a rest frame with respect to ${\rm SO}(3,1)$ in which it is equal to the $\rho_0=1$ circular hyperboloid. Because the standard texture is invariant under the Big Lorentz group, in this new frame ${\bf n}_0$ is unchanged. Now make the little ${\rm SO}(2,1)$ Lorentz transformation \eqref{lt} with $\rho_0=1$ then transform the hyperboloid and its new texture back to the original tilted, elliptical hyperboloid region $H$. We replace the bend-free interior of any shifted, tilted hyperboloid with a boosted bend-free texture. Finally, let the nematic strain tensor $C_{ij}\equiv \partial_i n_j$ \cite{epluribus} be ${\bf C}^{\hbox{\scriptsize int}}$ on the interior of $H$ and ${\bf C}^{\hbox{\scriptsize ext}}$ on the exterior.  Then this construction ensures that ${\bf C}^{\hbox{\scriptsize int}} - {\bf C}^{\hbox{\scriptsize ext}}$ is a rank-one matrix: the Hadamard jump condition is satisfied \cite{rankone}.

This process naturally suggests the formation of diabolo-shaped, geodesic domains -- packings of hyperboloids containing geodesic textures all matching on the boundary. We show here that such constructions can lower the free energy of the texture ${\bf n}_0$. In the presence of saddle-splay, the geodesic texture ${\bf n}_0$ is energetically favoured over the standard cholesteric in small diabolo domains~\cite{chaturvedi2020gnomonious}, but in larger systems the cholesteric texture will win out, since the twist decays to zero away from the origin. As the radius of the diabolo domains grows, we may therefore expect it to split into subdomains to regulate twist, which we expect to be hexagonally ordered; this is the structure shown in Figure~\ref{fig:energy}. By analogy, consider the double-twist tube in the blue phase.  A single tube can be dilated to fill space if the twist drops off to zero, precisely as it does in the global texture \eqref{ugh}.  But then the chiral term in the Frank free-energy does not enjoy the constant twist afforded by the standard single-twist cholesteric texture.  However, by creating a regular array of finite double-twist tubes, each tube enjoys a chiral energy close to that of the single-twist texture.  The saddle-splay compensates and the balance leads to the blue phase.  In our case, the chirality energy is also lowered by breaking the global gnomonic texture into diabolos.  Fortuituously, the Fitzgerald contraction associated with the little boosts reduce the splay as well by compressing the circular hyperboloids into elliptical hyperboloids, keeping the major axis fixed.

We construct a hexagonal domain as follows: Start with the skew domain \eqref{ugh} inside the hyperboloid $(x^2+y^2)/\rho^2 - z^2=1$, with some fixed vertical extent (for the calculations we take $|z|\leq2$). In the $xy$-plane, this domain is a circle of radius $\rho$. We take six circles of radius $\rho/6$ inside this domain, arranged hexagonally. Following the integral curves of ${\bf n}_0$ leads to six hyperboloids packed inside the original $\rho$ hyperboloid, shown in Figure~\ref{fig:energy}~(top left). We then apply the drill and fill construction: we remove the texture inside each of these hyperboloids and replace it with a boosted texture that matches on the boundary. In particular, we choose the `rest-frame' texture inside each of the six hyperboloids, so that the director field points parallel to the center line. The result is a texture inside the $\rho$ hyperboloid with six subdomains. While our construction ensures that the director field remains continuous, it is no longer differentiable and contains kinks (top center of Figure~\ref{fig:energy}) which contribute a boundary energy to the system proportional to the area, reminiscent of how FCDs are sewn together in a smectic texture.

To see if the hexagonal domains are preferable to the single domain, we compare their Frank free energies. In principle we must compute the splay, twist, bend and saddle-splay -- however the bend vanishes for both configurations. More subtly the saddle-splay, while not zero, is the {\sl same} in both cases. On each $z$ slice of a subdomain, the director field, thought of as a unit vector field, sweeps out an area on $\mathbb{S}^2$, related to saddle-splay~\cite{thegloriouskamienreview}. This area is invariant under the little Lorentz boosts, since the boundary hyperboloid does not change. Consequently the saddle-splay is the same for both the single domain and the hexagonal structure (or indeed for any set of subdomains). We therefore need only estimate splay and twist for the two configurations. There is an additional contribution from the domain interfaces which we do not estimate (though will be proportional to the boundary area at first order). Finally, we we take $q_0=1/2$ (following~\cite{chaturvedi2020gnomonious})\footnote{For the standard skew domain, the optimal value of $q_0$ is given by $3(1+ \log(1+\rho^2)/\rho^2)/(3+z_0^2)$, where $z_0$ is the vertical extent of the hyperboloid. For $z_0=2$ this ranges between $6/7$ for $\rho=0$, tending to $3/7$ as $\rho \to \infty$. $q_0=1/2$ falls within this range.}. The difference in splay and twist between then two textures is shown in Figure~\ref{fig:energy}~(bottom left), the hexagonal subdomains reduce the splay deformation, and there is a characteristic radius $\rho ~\sim \pi/2$ at which the reduction in twist is greatest, which sets the size.  When the contributions from the domain walls are sufficiently small, hexagonal packing of geodesic domains acts to lower the free energy.
 
\begin{figure}
\begin{center}
\includegraphics[scale=1]{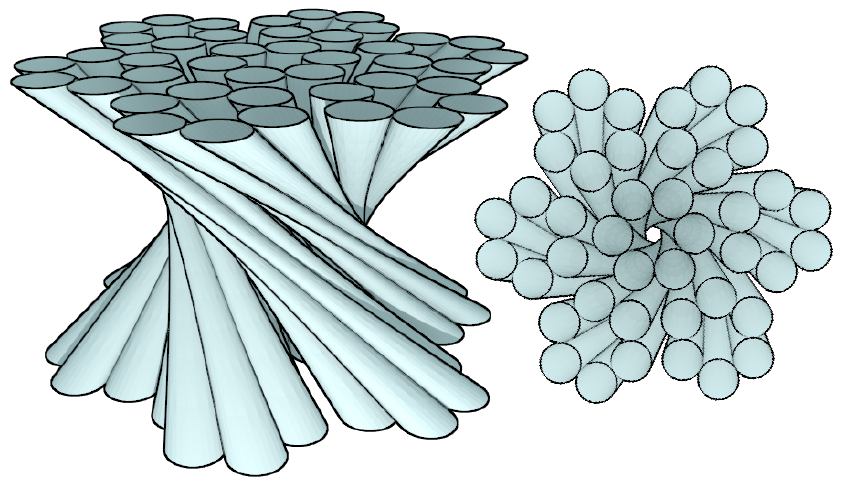}
\caption{Possible hierarchical domain structure, obtained by further splitting each hexagonal domain in Figure~\ref{fig:energy}. Such structures may become favourable as the outer domain size increases.}
\label{fig:hier}
\end{center}
\end{figure}

As the hyperbolic domain grows further, it may be energetically preferable for each sub-domain to yet again split, forming a hierarchical structure. A possibility is shown in Figure~\ref{fig:hier}. While this may be a crude mechanism to control the size of self-limited structures \cite{new1,new2}, there are a number of additional possibilities. For example, one could stack these structures, forming something akin to a hexagonal columnar phase or a moir\'e phase \cite{moire}. Though this would introduce bend defects it would lower the ever-growing splay of tall diabolic domains \cite{new6}. Another possibility is to construct an Apollonian packing of circles in the $z=0$ plane, leading to an intricate structure of skew domains. 

So far we have considered a restricted class of skew domains -- unnormalized, the director fields are quadratic in Cartesian coordinates. General bend-free textures have larger freedom, for example the local spacing between hyperboloids may be specified by an arbitrary function~\cite{chaturvedi2020gnomonious}. Moving to this larger class of textures leads to a more complex space of invariant surfaces, which are not hyperboloids. Results on the classification of geodesic fibrations of $\mathbb{R}^3$~\cite{salvai2009global,harrison2016skew,gluck1983great} suggest that this larger class of textures is topologically equivalent to the skew domains we study, and we do not anticipate any loss in qualitative power of our theory to describe diabolic domains in physical systems.

\section*{Lens domains: superluminal textures}

\begin{figure*}
\begin{center}
\includegraphics[scale=0.8]{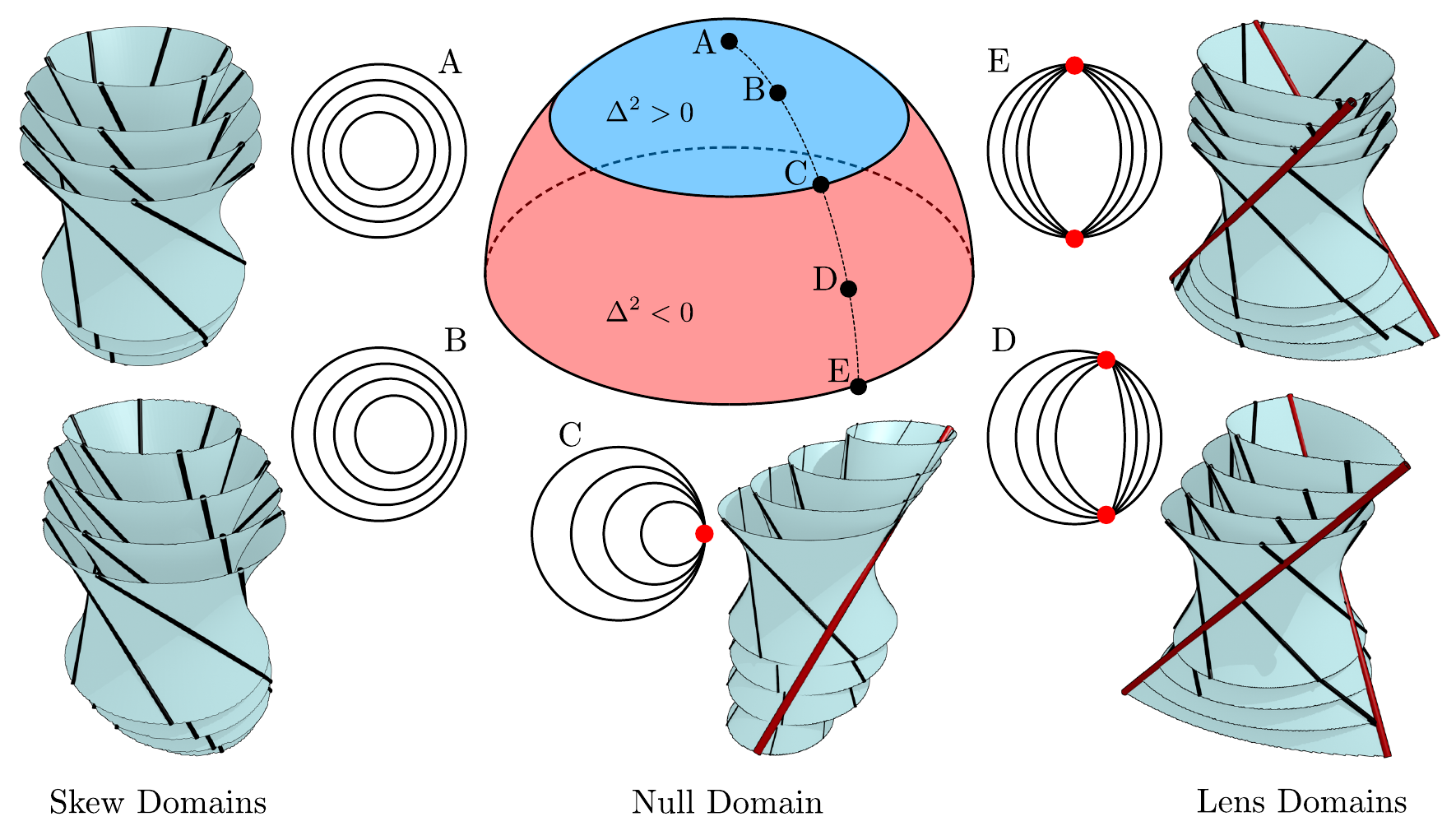}
\caption{Bend-free textures filling the hyperboloid $x^2+y^2-z^2=1$, parameterised by a unit vector ${\bf u}$ on a hemisphere ($\mathbb{R}{\bf P}^2$). Textures are classified according to the sign of the interval $\Delta^2 = u_z^2-u_x^2-u_y^2$. Within each class textures are equivalent up to Lorentz transformation and rotation. A, B are skew domains, C a null domain and D, E lens domains. Null and lens domains possess defect lines (red). A-E: Illustration of textures, using the concentric hyperboloids of Figure~\ref{fig:standard} under the sesquilinear transformation. Hyperboloids shown in blue with integral curves of ${\bf n}$ (black) and defect lines (red). The accompanying circles show the intersection with the $z=0$ plane.}
\label{fig:THELIST}
\end{center}
\end{figure*}

Applying the little Lorentz boost to ${\bf v}_0$ (taking $\rho_0=1$) yields
\begin{equation}\label{texture}
{\bf v} =  \gamma {\bf v}_0 + \beta \gamma \left [ 1-x^2, z  - x y, y - x z \right ]  = \gamma {\bf v}_0 + \beta \gamma {\bf v}_x.
\end{equation}
To obtain the director field we normalize, so $\gamma$ drops out, and the resulting texture is bend-free for {\sl all} values of $\beta$.  We may now choose to take $\beta \geq 1$, a superluminal boost! These superluminal textures cannot be obtained from ${\bf n}_0$ by ${\rm SO}(2,1)$ boosts and are topologically distinct from skew textures. They contain defects, which are straight lines on the $\rho_0=1$ hyperboloid. Recall that hyperboloids are {\sl doubly} ruled surfaces -- they contain a set of left-handed and right-handed straight lines. If the integral curves of ${\bf n}$ are left-handed rules then the defect lines are right-handed (and {\it vice versa}). For $\beta\geq1$ the defects lines, parameterized by $t\in\mathbb{R}$, are
\begin{equation}
{\bf d}_{\scriptscriptstyle\pm}(t)=\beta^{-1}\left (t  \pm \sqrt{\beta^2-1} ,\pm t  \sqrt{\beta^2-1} -1 , t \beta \right ),
\end{equation} 
For $\beta=1$ there is a single defect line with an index $2$ zero, while for $\beta>1$, the two defect lines each have index $1$. For reasons that will become clear below (see Figure~\ref{fig:THELIST}), we call the textures with $\beta=1$ null (or null domains) and for $\beta>1$ lens textures (or lens domains). These null and lens textures fit into our theory of domain packing. The interior of a hyperboloid domain may be either a skew, null, or lens domain. If, however, null or lens textures are used in the drill and fill procedure while keeping the exterior in skew form, then the defect profiles become $+1$ and $+1/2$ respectively, allowed by the nematic symmetry. We note that in chiral systems, lens domains are unlikely to be preferred as they do not have a single sign of ${\bf n} \cdot \nabla \times {\bf n}$, but may be observed as transients in coarsening \footnote{
This can be understood by writing the bend-free structures in a different way. A Killing vector field, ${\bf k}$, satisfies ${\bf k} \cdot {\bf n} = \textrm{const}$ if the integral curves of ${\bf n}$ are geodesics. It follows that the cross product of two Killing vector fields ${\bf k}_1$ and ${\bf k}_2$ defines a vector field whose integral curves are geodesics. In Euclidean $\mathbb{R}^3$ all Killing vector fields are of the form ${\bf k} = {\bf a} + {\bf r} \times {\bf b}$, for constant vectors ${\bf a}$, ${\bf b}$. By inspection one can see that the textures \eqref{eq:bt} with $u_x=0$ may be written as ${\bf k}_1 \times {\bf k}_2$ with ${\bf a}_1={\bf b}_1=\hat x$ and ${\bf a}_2=u_z\hat y - u_y\hat z$ and ${\bf b}_2=u_z\hat y + u_y\hat z$.
For $u_z>u_y$, both ${\bf k}_1$ and ${\bf k_2}$ define right-handed screw symmetries. For $u_y>u_z$, however, ${\bf k}_1$ is right-handed and ${\bf k}_2$ left-handed. This leads to the mixed handedness of the lens domains. }

Removing the subluminal constraint allows us to substitute $\gamma \to u_z$ and $\beta \gamma \to u_x$ in \eqref{texture}, where now $u_z$ and $u_x$ may take any values. We may also boost in any direction (not just $x$), and doing so (again with $\rho_0=1$) yields a texture which may be written as a linear combination
\begin{equation} \label{eq:bt}
{\bf v} = u_z {\bf v}_0 + u_x {\bf v}_x + u_y {\bf v}_y,
\end{equation}
with ${\bf v}_y$ the corresponding vector field obtained by a boost along $y$. As before, ${\bf n}$ is found by normalizing. In general we may write any texture whose integral curves are tangent to the $\rho_0=1$ hyperboloid in terms of the vector ${\bf u} = (u_x,u_y,u_z)$, and we may take $| {\bf u}|^2=1$. The vector ${\bf u}$ is equal to ${\bf n}$ at the origin and the `rest-frame' texture \eqref{ugh} corresponds to ${\bf u} = (0,0,1)$. The type of texture is then given by the sign of the interval $\Delta^2 = u_z^2-u_x^2-u_y^2$. Skew textures correspond to time-like intervals ($\Delta^2>0$), null textures to null intervals ($\Delta^2=0$), and lens textures to space-like intervals ($\Delta^2<0$). Since we are interested in director fields which are normalized and possess the nematic symmetry ${\bf n} \sim - {\bf n}$, we may take $u_z>0$. In this way, the set of textures filling a hyperboloid can be drawn as a hemisphere (a copy of $\mathbb{R}{\bf P}^2$), illustrated in Figure~\ref{fig:THELIST}.  This hemisphere of interior textures all match the boundary hyperboloid $\rho_0=1$ hyperboloid.  Since we may take any hyperboloid to any other via the ${\rm SO}(3,1)$ Big Lorentz group we can insert this whole hemisphere of fillings into {\sl any} of the infinite hyperboloids hiding in \eqref{ugh}.

Introducing the superluminal lens domains algebraically in this manner does not give any hint as to their geometric origins. How can we understand the geometry of the lens domains? As in the case of ${\bf n}_0$ we understand them in terms of a foliation of hyperboloids in space. But which hyperboloids? The little Lorentz transformation of the concentric hyperboloids changes the circular cross sections to concentric ellipses (Fitzgerald contraction), and indeed almost all hyperboloids in \eqref{aform} have an elliptic cross-section upon boosting. However, it turns out that there is a {\sl single} family of circles in the $xy$-plane that, under boost, remain circles. We choose boosts that preserve the $\rho_0=1$ hyperboloid and use the family $a_\mu = (\gamma a_0, -\gamma \beta a_0, 0,a_3)$ in \eqref{aform}. Following the convention in \eqref{eq:bt}, we rewrite $\gamma = u_z$ and $\gamma \beta= u_x$, realizing the hyperboloids:
\begin{eqnarray} \label{hyp0}
0&=& a_0 \left[ u_z  (1+x^2+y^2+z^2) + 2 u_x  (y-xz ) \right]\\ &&\qquad+a_3(x^2+y^2-z^2-1).\nonumber
\end{eqnarray}
Again, there is nothing prohibiting us from taking $u_x^2>u_z^2$, allowing the family of hyperboloids to describe superluminal textures. In the next section we show how this family arises naturally by considering transformations not in $\mathbb{R}^3$, but in $\mathbb{C}^2$. For now we describe them directly. First observe that the $a_3$ term does not depend on the boost -- it corresponds to the hyperboloid invariant under the little Lorentz transformation and, for $\beta = 0$, the family reduces to the concentric family centered on the $z$ axis. For $\beta \neq 0$, we find a family of concentric hyperboloids shifted in $y$ (recall we boosted in $x$, this is due to the skew nature of the texture), and tilted along $x$. Shown in Figure~\ref{fig:THELIST}(A-B). As $\beta$ approaches 1 (or $|u_x|$ approaches $|u_z|$) these hyperboloids begin to touch along a line Figure~\ref{fig:THELIST}(C). A shock forms, directly analogous to the Doppler effect. This shock is the single defect line in the null texture. As $\beta >1 $ the shock splits into two defects and the hyperboloids are no longer concentric -- instead they all pass through the two defect lines Figure~\ref{fig:THELIST}(D-E). Restricted to the interior of the $\rho_0=1$ hyperboloid, the super-luminal textures appear as nested lenses -- hence `lens domain'. \eqref{hyp0} defines a hyperboloid only if $a_3^2 \geq a_0^2 (u_z^2-u_x^2)$, in the super-luminal case $u_x>u_z$ and the equality is valid for all $a_0,a_3$, reflecting the change in topology of the textures. 
\section*{Projective transformations of geodesic domains}
Are there bend-free textures we may obtain from \eqref{ugh} other than by applying little Lorentz transformations? The most general transformation of $\mathbb{R}^3$ preserving zero-bend is fractional-linear in $(x,y,z)$, or equivalently linear in $(x,y,z,w)$. As before, we pick the subset of transformations that preserve a chosen hyperboloid, with the intent to apply the drill and fill construction. The group of real linear transformations of $(x,y,z,w)$ preserving the hyperboloid $x^2+y^2-z^2-w^2=0$ is $\textrm{SO}(2,2)$. Moreover, ${\rm SO}(2,2)$ is locally isomorphic to  ${\rm SU}(1,1) \times {\rm SU}(1,1)$.\footnote{Consider the identification
\begin{equation}
 (x,y,z,w) \to \left[\begin{matrix} 
x+z & y+w \\
w-y & x-z
\end{matrix}\right] = {\bf m},
\end{equation}
 $\det{\bf m} = x^2+y^2-z^2-w^2$ is invariant under ${\bf m}\rightarrow {\bf L}{\bf m}{\bf R}$ where ${\bf L},{\bf R}\in {\rm Sl}(2,\mathbb{R})$ and thus $\rm{SO}(2,2)\cong {\rm SL}(2,\mathbb{R})\times {\rm SL}(2,\mathbb{R})/{\pm I}$.  Representing an element of $\rm{SU}(1,1)$ in terms of two complex parameters $\alpha$ and $\delta$ with $\vert\alpha\vert^2-\vert\delta\vert^2=1$ as in \eqref{su11t}, we have
\begin{equation}
\left[\begin{matrix} 
{\rm Re}(\alpha+\delta) & {\rm Im}(\alpha-\delta) \\
-{\rm Im}(\alpha+\delta) & {\rm Re}(\alpha-\delta)
\end{matrix}\right] \in {\rm SL}(2,\mathbb{R}),
\end{equation}
establishing ${\rm SU}(1,1)\cong {\rm SL}(2,\mathbb{R})$.  Distinct skew textures are found by acting on the Hopf fibration with elements of ${\rm SU}(1,1)_R$. In this group there is a further $U(1)$ symmetry of the Hopf fibration, corresponding to $\alpha= e^{i \phi}$, $\delta = 0$. The space of skew textures is  ${\rm SU}(1,1)/{\rm U}(1) \cong \mathbb{H}$, the Hyperbolic plane. By the hyperboloid model, this is equivalent to time-like vectors in $M^3$ up to scaling. Finally, the groups are related via
\begin{equation}
\textrm{Big}={\rm SO}(3,1) \supset \textrm{SU}(1,1)_L \hookrightarrow \overline{{\rm SO}}(2,2) \twoheadrightarrow            \textrm{SU(1,1)}_R   \cong {\rm SO}(2,1)= \textrm{little}\nonumber
\end{equation}
where $\overline{{\rm SO}}(2,2)$ is the double cover of $\rm{SO}(2,2)$.
 }

The two copies of ${\rm SU}(1,1)$, which we label left ($L$) and right ($R$), are each determined by a pair of complex numbers $\alpha$, $\delta \in \mathbb{C}$ with $|\alpha|^2-|\delta|^2=1$. These act on the complex vector ${\bf Z}$ as
\begin{equation}\label{su11t}
\begin{bmatrix}
z_1 \\ z_2 \end{bmatrix} \mapsto 
\begin{bmatrix}
\alpha_{\scriptscriptstyle L} z_1+ \delta_{\scriptscriptstyle L} z_2 \\   \overline{\alpha}_{\scriptscriptstyle L} z_2 +\overline{\delta}_{\scriptscriptstyle L} z_1 \end{bmatrix}, \quad \begin{bmatrix}
z_1 \\ z_2 \end{bmatrix} \mapsto \begin{bmatrix}
\alpha_{\scriptscriptstyle R} z_1+ \overline{\delta}_{\scriptscriptstyle R} \overline{z}_2 \\ \alpha_{\scriptscriptstyle R} z_2 +\overline{\delta}_{\scriptscriptstyle R} \overline{z}_1  \end{bmatrix}.
\end{equation}
We see that ${\rm SU}(1,1)_{L}$ acts via complex-linear transformations of ${\bf Z}$ and is thus a subgroup of ${\rm SL}(2,{\mathbb C}) \cong {\rm SO}(3,1)$, {\it i.e.} the Big Lorentz group, which gives the (hidden) symmetries of the texture \eqref{ugh}. It follows that only ${\rm SU}(1,1)_R$, which acts sesquilinearly, may give non-trivial transformations of \eqref{ugh} used in the drill and fill construction.

How does the little Lorentz group ${\rm SO}(2,1)$ relate to these groups? ${\rm SO}(2,1)$ is the subgroup of $\rm{SO}(2,2)$ that leaves $w$ invariant. We may write it in terms of ${\rm SU}(1,1)_L \times {\rm SU}(1,1)_R$ with the constraints $\alpha_{\scriptscriptstyle L} = \alpha_{\scriptscriptstyle R}$ and  $\delta_{\scriptscriptstyle L} = \overline{\delta}_{\scriptscriptstyle R}$.
Because ${\rm SU}(1,1)_L$ does not change the basic texture \eqref{ugh} we can view the $\rm{SO}(2,2)$ transformations as either boosts in $\mathbb{R}^3$ using the little Lorentz group or as                      ${\rm SU}(1,1)_R$ transformations on ${\bf Z}$. This reflects the isomorphism ${\rm SU}(1,1) \cong {\rm SO}(2,1)$. Furthermore, we find that our classification in Figure~\ref{fig:THELIST} is complete. As an example, the  ${\rm SO}(2,1)$ boost with $\rho_0=1$ in \eqref{lt} gives the same transformation of \eqref{ugh} as the ${\rm SU}(1,1)_R$ map
\begin{equation} \label{trr}
\begin{bmatrix}
z_1 \\ z_2 \end{bmatrix} \mapsto
\begin{bmatrix}
z_1 \cosh \xi - \overline{z}_2 \sinh \xi  \\  z_2 \cosh \xi - \overline{z}_1\sinh \xi  \end{bmatrix},
\end{equation}
where $\tanh 2\xi=\beta$ (note the standard spinor factor of 2). While the little Lorentz group $\rm{SO}(2,1)$ and ${\rm SU}(1,1)_R$ transformations treat the texture the same, they transform the hyperboloids \eqref{qform} differently. Indeed one may check that the hyperboloids \eqref{hyp0} are precisely the transformation of the concentric hyperboloids under ${\rm SU}(1,1)_R$ -- their geometric naturality is revealed. Representing transformations of \eqref{ugh} using ${\rm SU}(1,1)_R$ also allows us to understand the superluminal boosts. The hyperboloid $|z_1|^2-|z_2|^2=0$  is invariant not only under ${\rm SO}(2,2)$, but also under transformations that {\sl reverse} the sign of $|z_1|^2-|z_2|^2$, these are superluminal (switch $\cosh$ and $\sinh$ in \eqref{trr}). The sub- and super- luminal boosts may be unified by complexifying $(x,y,z,w)$, but we need not develop that perspective here. 

\section*{Lens domains and Milnor fibrations}

Before concluding we interpret the geometry of the lens domains.  The prototypical skew domain \eqref{ugh} is related to the Hopf fibration on $\mathbb{S}^3$ by gnomonic projection. There is, similarly, a prototypical lens domain expressed through \eqref{texture} with $u_z=0$ and $u_x=1$. These textures are also natural on $\mathbb{S}^3$: they are related to the Milnor fibration of the Hopf link. Consider the family of hyperboloids in \eqref{hyp0} with $u_x=1$, $u_z=0$. Since $|u_x|>|u_z|$, we have a hyperboloid for all $a_0$, $a_3$. We may therefore take $a_0^2+a_3^2=1$ and write the family of hyperboloids as
\begin{equation}
{\rm Re}\left [ e^{-i \phi} (z_1 - i \overline{z}_2)( \overline{z}_1 - i z_2 ) \right ] = {\rm Re}\left [ e^{-i \phi} f \right ] = 0,
\end{equation}
for ${\rm atan}(a_0/a_3) = \phi \in [0, 2\pi)$. 
The zeros of $f$ are given by the equations $z_1 = \pm i \overline{z}_2$, which on $\mathbb{S}^3$ correspond to two fibers of the right-handed Hopf fibration, forming a pair of linked great circles -- a Hopf link. In $\mathbb{R}^3$ these are the defect lines of the texture. Following the work of Milnor~\cite{milnor1968singular}, the level sets of $\phi  = \textrm{Arg}\, f$ are Seifert surfaces $\Sigma_\phi$ for this Hopf link, and give an open book decomposition of $\mathbb{S}^3$. Each $\Sigma_\phi$ is a piece of a Clifford torus in $\mathbb{S}^3$, and so has two rulings by arcs of great circles. By inspection, one ruling corresponds to the great circles traced out by $(z_1, z_2) \mapsto (e^{i \theta} z_1, e^{-i \theta} z_2)$, the right-handed Hopf fibration (two of these circles correspond to the zeros of $f$ -- the defect lines). Now fill $
\mathbb{S}^3$ by drawing the {\sl other} set of rulings on each $\Sigma_\phi$. The resulting set of great circles on $\mathbb{S}^3$ all intersect the two defect lines, and the gnomonic projection of this structure down to $\mathbb{R}^3$ gives the lens domain.

\section*{Conclusion}

We have introduced a new construction of continuous bend-free textures comprised of diabolic domains. We note a very close relation with the structure of FCDs.  Recall that equal spacing of smectic layers implies $({\bf n}\cdot\nabla){\bf n}=0$ \cite{lens1,lens2,klemanlens,achard} but, in the case of smectics the twist must vanish identically.  In our study the twist is almost-everywhere non-zero (in the case of skew domains, it is strictly non-zero).   Just as a complex smectic texture can be built out of FCDs to guaranty constant spacing, the diabolical domains can be employed to find low-energy chiral nematic complexions with zero bend.  It is amusing that both the diabolical and focal conic constructions can be interpreted from the perspective of Lorentz transformations \cite{poincare}: the hint of an even deeper connection intrigues.  Whether generalizations to higher-dimensional fibrations \cite{hopf} exist and whether they have a relevant physical realization remains an open question.

Where might these textures be detected?  Highly chiral systems with large $K_3$ and an appreciable $K_{24}$ are candidates.  Studies of the liquid crystalline phases of nucleosome core particles have observed diabolo patterns akin to those in Fig. 1 \cite{LL}.  Recently, there have been studies of achiral chromonic liquid crystals \cite{mm,jj} in confinement.  Perhaps, made chiral and set free, these materials would be a good starting point.  Likewise, simulations of achiral, bent-core, rigid molecules have observed the spontaneous formation of diabolic domains \cite{new7,new8} -- perhaps further studies would uncover the hidden hyperboloids presented here and find a numerical manifestation of the drill and fill construction. How would they be detected?  A tell-tale sign of these structures is that, as we have demonstrated from the action of ${\rm SL}(2,\mathbb{C})$, if we start with the standard texture ${\bf n}_0$ then all the diabolical domains intersect the $z=0$ plane as {\sl circles}.  Cross sections along any other plane will cut through domains in ellipses.  In addition to the fact that our structures are macroscopically chiral, the elliptical cross sections here would be different than the polygonal textures in smectics \cite{friedel}: there is no ``law of corresponding cones,'' and so we might expect less order in the diabolical ellipses.  
\begin{acknowledgements}
It is a pleasure to acknowledge stimulating discussions with G.P. Alexander, N. Chaturvedi, G.M. Grason, J.H. Hannay, C.D. Modes, and L. Tran.  This work was supported by a Simons Investigator grant from the
Simons Foundation to R.D.K. 
\end{acknowledgements}


\begin{thebibliography}{21}
\providecommand{\natexlab}[1]{#1}
\providecommand{\url}[1]{\texttt{#1}}
\expandafter\ifx\csname urlstyle\endcsname\relax
  \providecommand{\doi}[1]{doi: #1}\else
  \providecommand{\doi}{doi: \begingroup \urlstyle{rm}\Url}\fi



\bibitem[Landau and Lifshits(1935)]{LanLif}
L.~Landau and E.~Lifshits.
\newblock On the theory of the dispersion of magnetic permeability in
  ferromagnetic bodies.
\newblock \emph{Phys. Z. Sowjetunion} {\bf 8}, 153--169 (1935).


\bibitem[Hirth and Lothe(1992)]{hirth1992theory}
J.P. Hirth, J. Lothe.
\newblock \emph{Theory of Dislocations}.
\newblock (Krieger Publishing Company, 1992).

\bibitem[ABRIKOSOV(1957)]{Ab}
A.~A. Abrikosov
\newblock On the magnetic properties of superconductors of the second group.
\newblock \emph{Sov. Phys. JETP} {\bf 5}, 1174--1182 (1957).


\bibitem[Hall et~al.(1956{\natexlab{a}})Hall, Vinen, and Shoenberg]{HV1}
H.~E. Hall, W.~F. Vinen, D. Shoenberg.
\newblock The rotation of liquid helium II i. experiments on the propagation of
  second sound in uniformly rotating helium ii.
\newblock \emph{Proc. R. Soc. Lond. A} {\bf 238}, 204--214 (1956).



\bibitem[Hall et~al.(1956{\natexlab{b}})Hall, Vinen, and Shoenberg]{HV2}
H.~E. Hall, W.~F. Vinen, D. Shoenberg
\newblock The rotation of liquid helium II ii. the theory of mutual friction in
  uniformly rotating helium ii.
\newblock \emph{Proc. R. Soc. Lond. A} {\bf 238}, 215--234 (1956).


\bibitem[Renn and Lubensky(1988)]{Renn:1988p2132}
S. Renn, T.~C. Lubensky.
\newblock Abrikosov dislocation lattice in a model of the cholesteric to
  smectic a transition.
\newblock \emph{Phys. Rev. A} {\bf 38}, 2132--2147 (1988).

\bibitem[{de Gennes}(1972)]{dG}
P.~G. {de Gennes}.
\newblock {An analogy between superconductors and smectics A}.
\newblock \emph{Solid State Commun.} {\bf 10},
  753--756 (1972).


\bibitem[Kl{\'e}man(1983)]{plw}
M. Kl{\'e}man.
\newblock \emph{Points, lines, and walls: in liquid crystals, magnetic systems,
  and various ordered media}.
\newblock (J. Wiley, Chichester ; New York, 1983).


\bibitem[Friedel(1922)]{friedel}
G.~Friedel.
\newblock Les {\'e}tats m{\'e}somorphes de la mati{\`e}re.
\newblock \emph{Ann. Phys. (Paris)} {\bf 18}, 273--474 (1922).

\bibitem[Kl{\'e}man and Lavrentovich(2000)]{grainb}
M.~Kl{\'e}man, O.~D. Lavrentovich.
\newblock Grain boundaries and the law of corresponding cones in smectics.
\newblock \emph{Eur. Phys. J. E} {\bf 2},  47--57 (2000).

\bibitem[Meyer et~al.(2009)Meyer, {Le Cunff}, Belloul, and Foyart]{flower1}
C.~Meyer, L.~{Le Cunff}, M.~Belloul, G.~Foyart.
\newblock {Focal Conic Stacking in Smectic A Liquid Crystals: Smectic Flower
  and Apollonius Tiling}.
\newblock \emph{Materials} {\bf 2}, 499--513 (2009)


\bibitem[Beller et~al.(2013)Beller, Gharbi, Honglawan, Stebe, Yang, and
  Kamien]{flower2}
D.~A. Beller, M.~A. Gharbi, A. Honglawan, K.~J. Stebe, S.
  Yang, R.~D. Kamien.
\newblock Focal conic flower textures at curved interfaces.
\newblock \emph{Phys. Rev. X} {\bf 3}, 041026 (2013).

\bibitem[Chaturvedi and Kamien(2020)]{chaturvedi2020gnomonious}
N. Chaturvedi, R.~D Kamien.
\newblock Gnomonious projections for bend-free textures: thoughts on the
  splay-twist phase.
\newblock \emph{Proc. R. Soc. A} {\bf 476}, 20190824 (2020).



\bibitem[Frank (1958)]{frank58}
F.~C. Frank. {\em I. Liquid crystals. On the theory of liquid crystals}, Faraday Discuss. {\bf 25}, 19--28 (1958).

\bibitem[Sethna et~al.(1983)Sethna, Wright, and Mermin]{sethna1983relieving}
J.~P. Sethna, D.~C. Wright, N.~D.~Mermin.
\newblock Relieving cholesteric frustration: the blue phase in a curved space.
\newblock \emph{Phys. Rev. Lett} {\bf 51}, 467 (1983).


\bibitem{new4}
I. Niv and E. Efrati.
\newblock \emph{Geometric frustration and compatibility conditions for two-dimensional director fields}.
\newblock \emph{Soft Matter} {\bf 14}, 424--431 (2018).


\bibitem{new3}
E.G. Virga.
\newblock \emph{Uniform distortions and generalized elasticity of liquid crystals}.
\newblock \emph{Phys. Rev. E} {\bf 100}, 052701 (2019).


\bibitem{new5}
J.F. Sadoc, R. Mosseri, and J.V. Selinger.
\newblock \emph{Liquid crystal director fields in three-dimensional non-Euclidean geometries}.
\newblock \emph{arXiv:2006.12668}.

\bibitem[Beller et~al.(2014)Beller, Machon, \ifmmode~\check{C}\else
  \v{C}\fi{}opar, Sussman, Alexander, Kamien, and Mosna]{epluribus}
D.~A. Beller, T. Machon, S.\ifmmode~\check{C}\else \v{C}\fi{}opar,
  D.~M. Sussman, G.~P. Alexander, R.~D. Kamien, R.~A.
  Mosna.
\newblock Geometry of the cholesteric phase.
\newblock \emph{Phys. Rev. X} {\bf 4}, 031050 (2014).

\bibitem[Ball and James(1987)]{rankone}
J.~M. Ball, R.~D. James.
\newblock Fine phase mixtures as minimizers of energy.
\newblock \emph{Arch. Ration. Mech. Anal.} {\bf 100}, 13--52, 1987.

\bibitem{thegloriouskamienreview}
R.D. Kamien.
\newblock \emph{The geometry of soft materials: a primer}.
\newblock \emph{Rev. Mod. Phys.} {\bf 74}, 953 (2002).
\bibitem{new1}
J.M. Miller, C. Joshi, P. Sharma, A. Baskaran, A. Baskaran, G.M. Grason, M.F. Hagan, and Z. Dogic.
\newblock \emph{Conformational switching of chiral colloidal rafts regulates raft-raft attractions and repulsions}.
\newblock \emph{Proc. Natl. Acad. Sci.} {\bf 116}, 15792--15801 (2019).

\bibitem{new2}
G.M. Grason.
\newblock \emph{Perspective: Geometrically frustrated assemblies}.
\newblock \emph{J. Chem. Phys.} {\bf 145}, 110901 (2016).

\bibitem[Kamien and Nelson(1996)]{moire}
R.~D. Kamien, D.~R. Nelson.
\newblock Defects in chiral columnar phases: Tilt-grain boundaries and iterated
  moir{\'e} maps.
\newblock \emph{Phys. Rev. E} {\bf 53}, 650 (1996).

\bibitem{new6}
J.M. Miller, D. Hall, J. Robaszewski, P. Sharma, M.F. Hagan, G.M. Grason, and Z. Dogic.
\newblock \emph{All twist and no bend makes raft edges splay: Spontaneous curvature of domain edges in colloidal membranes}.
\newblock \emph{arXiv:1908.09966}.



\bibitem[Salvai(2009)]{salvai2009global}
M. Salvai.
\newblock Global smooth fibrations of $\mathbb{R}^3$ by oriented lines.
\newblock \emph{B. Lond. Math. Soc.} {\bf 41}. 155--163 (2009).

\bibitem[Harrison(2016)]{harrison2016skew}
M. Harrison.
\newblock Skew flat fibrations.
\newblock \emph{Math. Z.} {\bf 282},
  203--221 (2016).

\bibitem[Gluck and Warner(1983)]{gluck1983great}
H. Gluck, F.~W Warner.
\newblock Great circle fibrations of the three-sphere.
\newblock \emph{Duke Math. J.} {\bf 50}, 107--132 (1983).

\bibitem[Milnor(1968)]{milnor1968singular}
John Milnor.
\newblock \emph{Singular points of complex hypersurfaces}.
\newblock Number~61. Princeton University Press, 1968.

\bibitem{klemanlens}
C. Blanc and M. Kl\'eman.
\newblock \emph{The confinement of smectics with a strong anchoring}.
\newblock \emph{Eur. Phys. J. E} {\bf 4}, 241--251 (2001).

\bibitem{achard}
M.-F. Achard, M. Kl\'eman, Yu.A. Natishin, and H.-T. Nguyen.
\newblock \emph{Liquid crystal helical ribbons as isometric textures}.
\newblock \emph{Eur. Phys. J. E} {\bf 16}, 37--47 (2005).

\bibitem{lens1}
C.D. Santangelo, V. Vitelli, R.D. Kamien, and D.R. Nelson.
\newblock \emph{Geometric Theory of Columnar Phases on Curved Substrates}.
\newblock \emph{Phys. Rev. Lett.} {\bf 99}, 017801 (2007).

\bibitem{lens2}
R.D. Kamien, D.R. Nelson, C.D. Santangelo, and V. Vitelli.
\newblock \emph{Extrinsic Curvature, Geometric Optics, and Lamellar Order on Curved Substrates}.
\newblock \emph{Phys. Rev. E} {\bf 80}, 051703 (2009).

\bibitem{poincare}
G.P. Alexander, B.G. Chen, E.A. Matsumoto, and R.D. Kamien.
\newblock \emph{The Power of Poincar\'e: Elucidating the Hidden Symmetries in Focal Conic Domains}.
\newblock \emph{Phys. Rev. Lett.} {\bf 104}, 257802 (2010).

\bibitem[Ovsienko and Tabachnikov(2016)]{hopf}
V. Ovsienko and S. Tabachnikov.
\newblock Hopf Fibrations and Hurwitz-Radon Numbers.
\newblock \emph{Math. Intelligencer} {\bf 38}, 11--18 (2016).

\bibitem{LL}
F. Livolant and A. Leforestier.
\newblock \emph{Chiral discotic columnar germs of nucleosome core particles|}.
\newblock \emph{Biophys. J.} {\bf 78}, 2716--2729 (2000).

\bibitem{mm}
K. Nayani, R. Chang, J. Fu, P.W. Ellis, A. Fernandez-Nieves, J.O. Park and M. Srinivasarao.
\newblock \emph{Spontaneous emergence of chirality in achiral lyotropic chromonic liquid crystals confined to cylinders}.
\newblock \emph{Nat. Comm} {\bf 6}, 8067 (2015).

\bibitem{jj}
J. Jeong, L. Kang, Z.S. Davidson, P.J. Collings, T.C. Lubensky, and A.G. Yodh.
\newblock \emph{Chiral structures from achiral liquid crystals in cylindrical capillaries}.
\newblock \emph{Proc. Natl. Acad. Sci} {\bf 112}, E1837--1844 (2015).



\bibitem{new7}
F. Yan, C.A. Hixson, and D.J. Earl.
\newblock \emph{Self-Assembled Chiral Superstructures Composed of Rigid Achiral Molecules
and Molecular Scale Chiral Induction by Dopants}.
\newblock \emph{Phys. Rev. Lett.} {\bf 101}, 157801 (2008).

\bibitem{new8}
F. Yan, C.A. Hixson, and D.J. Earl.
\newblock \emph{Computer simulations of linear rigid particles that form chiral superstructures
and tilted smectic phases}.
\newblock \emph{Soft Matter} {\bf 5}, 4477--4483 (2009).



\end{thebibliography}
\end{document}